\documentclass{aa}
 
\newcommand\Tstrut{\rule{0pt}{2ex}}         
\newcommand\Bstrut{\rule[-0.9ex]{0pt}{0pt}}   
\usepackage{graphicx}
\usepackage{txfonts}
\begin{document}

   \title{Astronomical image time series classification using CONVolutional attENTION (ConvEntion) }

   \author{Anass Bairouk
          \inst{1},
          Marc Chaumont\inst{1,3}, Dominique Fouchez\inst{2}, \\ Jerome Paquet\inst{4, 5}, Frédéric Comby\inst{1}
          , Julian Bautista\inst{2}
          }

   \institute{Laboratory of Computer Science, Robotics and Microelectronics of Montpellier, University of Montpellier, 34095 Montpellier, France\\
              \email{anass.bairouk@lirmm.fr}
         \and
            Aix Marseille Univ, CNRS/IN2P3, Centre of Particle Physics of Marseilles, 13009  Marseille, France
        \and
              University of Nimes, 30021 Nîmes, France
        \and
             Groupe AMIS, Paul Valéry University Montpellier 3, 34090 Montpellier, 
               France
         \and
              Land, Environment, Remote Sensing and Spatial Information - UMR TETIS, INRAE/CIRAD/CNRS, 34000 Montpellier, France
             }


 
  \abstract
   {}
   {
   The treatment of astronomical image time series has won increasing attention in recent years.
Indeed, numerous surveys following up on transient objects are in progress or under construction, such as the Vera Rubin Observatory Legacy Survey for Space and Time (LSST), which is poised to produce huge amounts of these time series.
The associated scientific topics are extensive, ranging from the study of objects in our galaxy to the observation of the most distant supernovae for measuring the expansion of the universe. With such a large amount of data available, the need for robust automatic tools to detect and classify celestial objects is growing steadily.
     
}
   {This study is based on the assumption that astronomical images contain more information than light curves. In this paper, we propose a novel approach based on deep learning for classifying different types of space objects directly using images. We named our approach ConvEntion, which stands for CONVolutional attENTION. It is based on convolutions and transformers, which are new approaches for the treatment of astronomical image time series. Our solution integrates spatio-temporal features and can be applied to various types of image datasets with any number of bands.}
   {In this work, we solved various problems the datasets tend to suffer from and we present new results for classifications using astronomical image time series with an increase in accuracy of 13\%, compared to state-of-the-art approaches that use image time series, and a 12\% increase, compared to approaches that use light curves.}
   {}

\titlerunning{ConvEntion}
\authorrunning{A. Bairouk et al}
   \maketitle
%
\section{Introduction}

 \begin{figure*}
  \includegraphics[width=18cm, height=9.161cm]{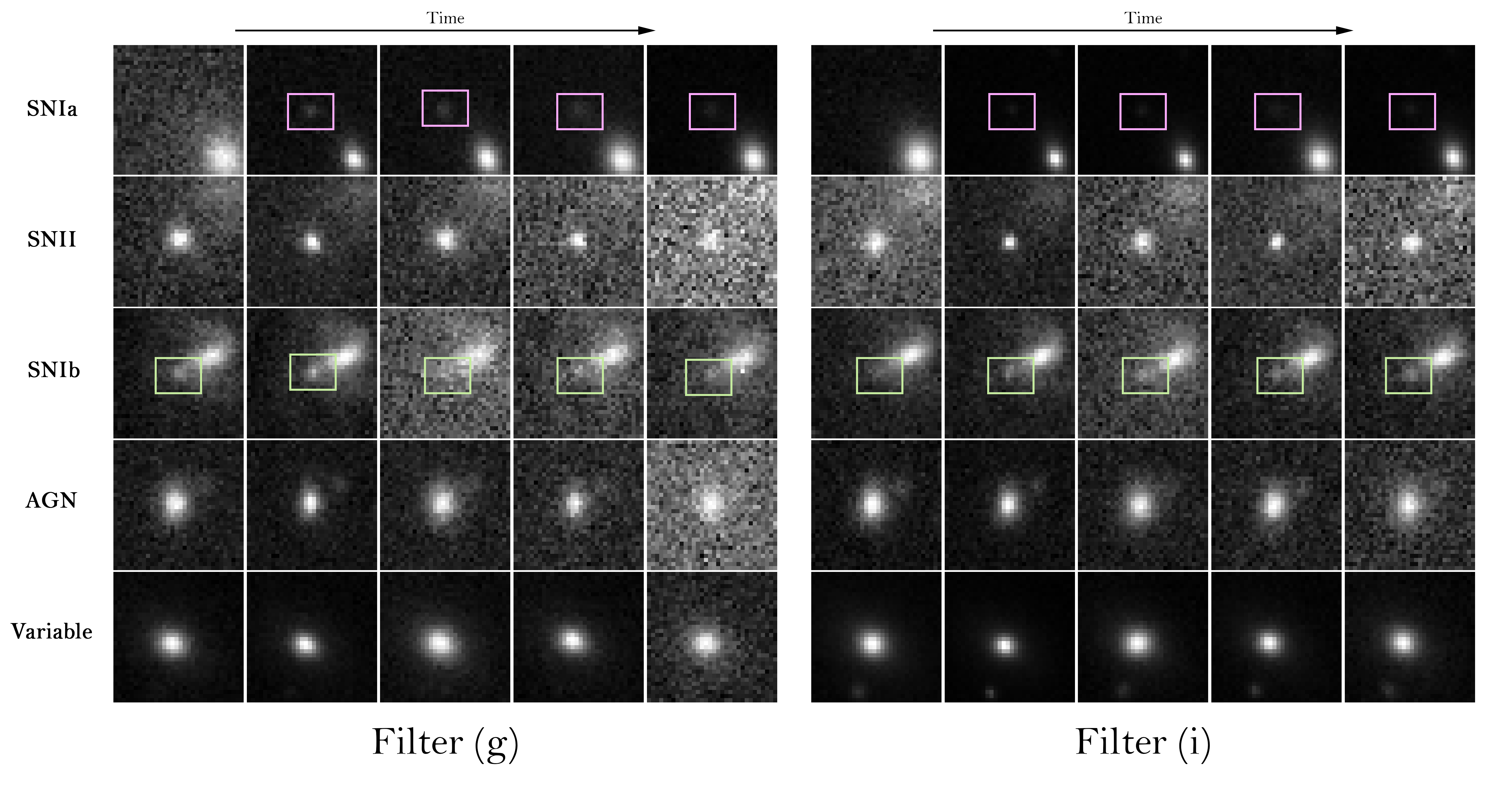}
  \caption{Sample of some objects present in our dataset. Each image in filter g/i corresponds to a different observation with the same filter.}
  \label{fig:imagedata}
\end{figure*}

The astronomical community  has been facing a considerable challenge in the last few years as tools for observing the universe continue to improve. Telescopes are becoming more powerful, with the capacity to observe a huge part of the universe and generate a massive amount of data. Processing and analyzing these data are very demanding steps in terms of their computational and human resource requirements. With the promises of The Vera Rubin Observatory Legacy Survey for Space and Time (LSST)  \citep{lsstser}, the field will see the discovery of 10 to 100 times more astronomical sources that fluctuate in the night sky. Some of these sources will be entirely new. LSST is prepared to alert the community to 10 million new objects per night, and these objects all need to be classified. There are many types of objects, including active galactic nuclei (AGNs), variables, cepheids, RR Lyrae, and supernovae. The latter stands the most important transient object for cosmology because increasingly large samples of Type Ia supernovae (SNe Ia) are being used to measure luminosity distances as a function of redshift in order to understand the origin of the acceleration of the expansion of the universe.

Traditionally, the classification of these objects goes through many processes in a complex pipeline. First, the preprocessing phase known as photometry is conducted on a series of images to extract the flux per band, each band corresponding to a passband-like color filter. The number of bands can vary, depending on the survey, for example SDSS survey \citep{sdss,sdss2,sdss1}  has five bands and the Catalina survey \citep{Drake_2011} has only one band.
Then,  a time series of brightness changes is generated over time, called the light curves. Afterwards, the light curve is fed to a machine-learning classifier to determine the class of the object. Among all the methods developed to perform such a classification,  \citet{b18} 
introduced a model called SuperNNova: a supernova photometric classification framework that uses a recurrent neural network (RNN) \citep{rnn,lstm,gru} to classify different types of supernovas such as SNIa, SNIb, SNIIP, and others using only light curves. 
The proposition yields good results because while Bayesian neural networks (BNN) are known to be robust to overfitting and can easily learn from small datasets, they are still significantly more complex than standard neural networks and computationally expensive.
\citet{b15} (winner of the photometric classification challenge PLAsTiCC \citep{b17, b16}) presented a model based on Gaussian process augmentation of the light curve and then train it on boosted decision tree classifier.  \citet{b19} created a deep architecture called PELICAN that accepts only light curves and redshifts as input. PELICAN can handle light curves with sparsity and irregular sampling. Some can choose to add more preprocessing before training a model. For instance, \citet{Scone} proposed a novel approach where they generated a 2D image heatmap from light curves using 2D Gaussian process regression, which they fed to convolutional neural networks to classify different types of supernovae.
The approach yields great results on PLAsTiCC data, with an accuracy of 99.73\% on the binary classification of SNIa and non-SNIa.
 The methods that use light curves for classification still have some limitations. In order to generate a light curve, we should correctly align  the two consecutive images and we must lower the quality of one of the two images to subtract them to get the flux, which could lead to a loss of information. 
Some dedicated algorithms called scene modeling can mitigate such issues on blended objects but are very demanding in terms of computer resources. Most importantly, the scene information, namely, the background of the transient object, is in general not taken into account in the classification.
Several recent works have proposed to eliminate the feature extraction and light curve phase and focus on classifying the objects using only images. \citet{b20} and \citet{b21} used a RNN to classify the sequences after passing the images through a CNN to extract the spatial features. They forwarded the output to the RNN (GRU/LSTM) to extract the temporal characteristics and classify the object, while \citep{b21} applied their model to only transient objects and \citet{b20} classified variables and transient. These two papers showed promising results for the astronomical image time series (AITS). Therefore, we followed the same path to improve the classification and also solve some challenges posed by AITS, which have not been tackled before.

In particular, image time series (ITS) classification has always been one of the challenging areas of deep learning. In addition to spatial characteristics, you also need to give importance to the temporal aspects, which makes traditional feed-forward networks ineffective. Due to the lack of research carried out on ITS in astronomy, we need to import new technics from other fields of research. Most of the research in ITS classification is done in two major domains: action recognition, where the goal is to classify the type of human action \citep{b4,b22}, and landscape classification using satellite images \citep{b5}. These two fields have covered many of the essential methods to handle ITS.
Then, RNN-based approaches use recurrent neural networks to manage the aspect of time in the classification. These approaches are split into two main categories. The first one handles the spatial features separately from the temporal features. \citet{b20} and \citet{b21} used precisely this method at the point when the CNN handles the spatial characteristics to pass it later to the RNN, which might be LSTM \citep{lstm} or GRU \citep{gru}. The second category combines convolution inside the RNN cell, thus maintaining the spatial structure of the input, which leads to extracting spatial-temporal features in the sequence. This method was first introduced by \citet{b4}. These authors demonstrated how to create an end-to-end trainable model using the convolutional LSTM (ConvLSTM). Experiments indicate that their ConvLSTM network regularly exceeds fully connected LSTM (FC-LSTM) in capturing Spatio-temporal correlations. Using satellite images, \citet{b5} proposed a new type of RNN called ConvSTAR,  which has fewer parameters than the LSTM and GRU.
Another way of achieving the classification of ITS is by using convolution neural networks.  \citet{b22} created a new 3D CNN model for action recognition. This model pulls features from spatial and temporal dimensions, collecting motion information contained in several consecutive frames. Meanwhile, some of the latest developments have abandoned convolutions and RNNs to replace them with only transformers.  \citet{swin} and \citet{ multiview} proposed an improved supervised transformer for image classification. On the other hand,  \citet{ibot} and \citet{beit} proposed more complex transformers that are self-supervised.

In this work, we develop a new deep learning transformer-based architecture to classify AITS. Unlike other works that separate spatial and temporal feature extraction, we combine these two steps by performing a spatio-temporal feature extraction in one step. It improves the capacity of the network to recognize the objects. We also propose a solution for the missing observations problem, which demonstrates a significant improvement in the accuracy of the model. To illustrate the performances of our model, we tested it with actual data from the SDSS survey  \citep{sdss,sdss2,sdss1}. In Section \ref{section:dataset}, we describe the dataset that we used in our work. Section \ref{section:methods} introduces our architecture ConvEntion and describes the role of each component of the model. In Section \ref{section:experiments}, we present the results of our work with some statistics about the performance and some comparisons with other architectures used for image time series classification. Finally, in Section \ref{section:conclusion}, we  present our conclusions and perspectives on this work.
\begin{figure*}[ht]
  \centering{\includegraphics[width=17cm, height=10.39cm]{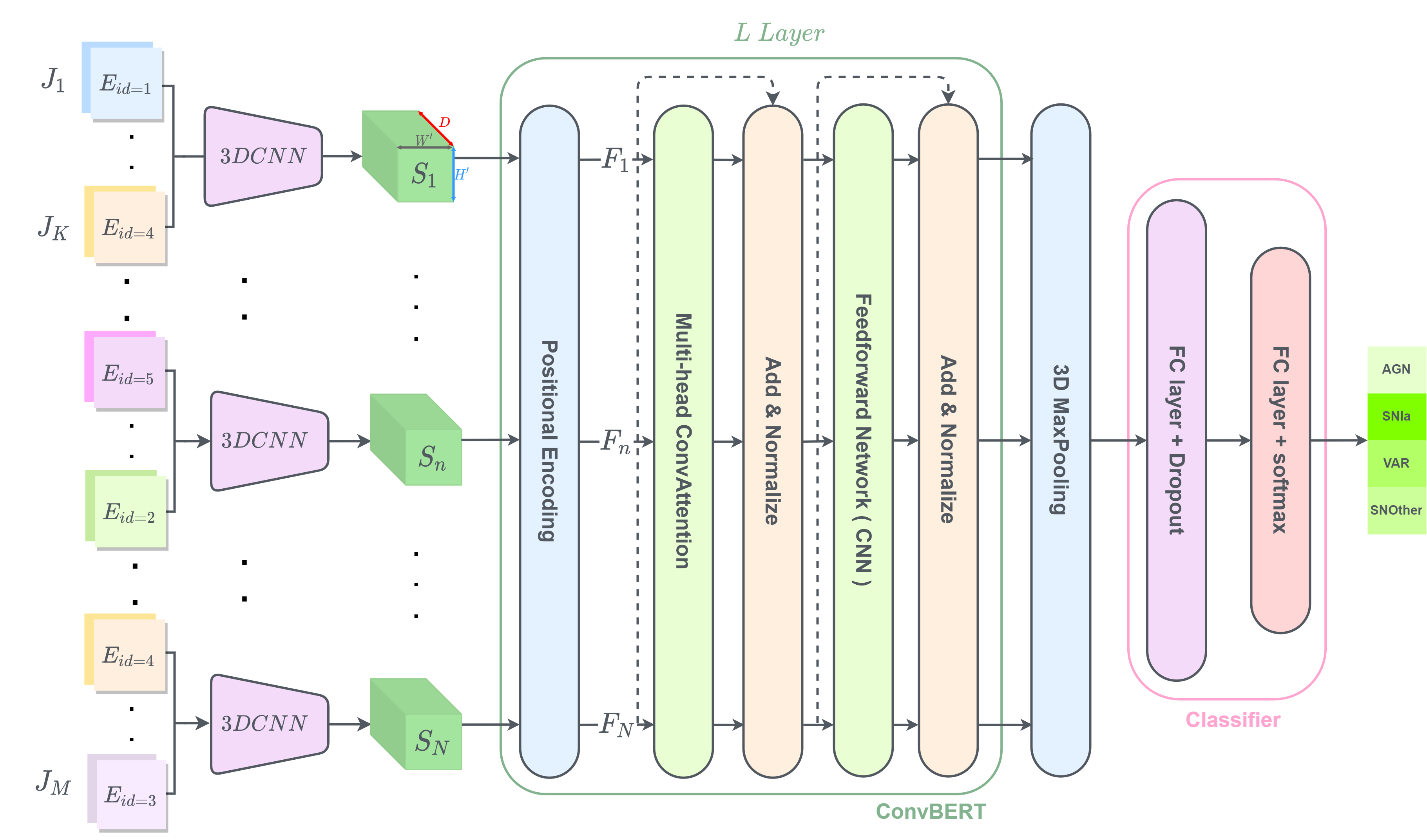}}
  \caption{General architecture of the ConvEntion network. The image time series are first rearranged to embed the band information. Then each 3DCNN is fed with a sub-sequence of $K$ inputs of the time series ${J} (\in \mathbb{R}^{M \times H \times W \times 2}$ for M elements of images of size HxW) to create the new downsized sequence ${S} (\in \mathbb{R}^{N \times H' \times W' \times D})$. $S$ is fed to the positional encoder in order to add the information about the position, which outputs ${F} (\in \mathbb{R}^{N \times H' \times W' \times D})$. Then $F$ is passed to ConvBERT which has $L$ layers. The 3D max-pooling is used to downsize the output of ConvBERT for the classifier}
  \label{fig:allarchi}
\end{figure*}

\section{Dataset}
\label{section:dataset}
\subsection{Database description}

The Sloan Digital Sky Survey (SDSS) \citep{sdss,sdss1} is a very ambitious and successful large-scale survey program using a dedicated 2.5-meter telescope at Apache Point Observatory, New Mexico, equipped with photometric and spectroscopic instruments that have released images, spectra, and catalog information for several hundred million celestial objects. The dataset used in this paper was  collected during the SDSS Supernova Survey \citep{sdss2}, one of three components (along with the Legacy and SEGUE surveys) of SDSS-II, a three-year extension of the original SDSS that operated from July 2005 to July 2008. The Supernova Survey is a time-domain survey, involving repeat imaging of the same region of the sky every other night, weather permitting. 

The images are obtained through five wide-band filters  \citep{sdss96}
named u', g', r', i', and z', simplified as u, g, r, i, and z in the following, which corresponds to an effective mid-point wavelength of u (365nm), g (475nm), r (658nm), i (806nm), and z (900nm).
The survey region observed repeatedly over three years is a 2.5-degree-wide stripe centered on the celestial equator in the Southern Galactic Cap that has been imaged numerous times in the last twenty years, allowing for the construction of a big image database for the discovery of new celestial objects. Most of the sources have included galactic variable stars, active galactic nuclei (AGN), supernovae (SNe), and other astronomical transients, all of which have been processed to generate multi-band (ugriz) light curves.
The imaging survey is reinforced by an extensive spectroscopic follow-up program that uses spectroscopic diagnostics to identify SNe and measure their redshifts. Light curves were evaluated during the survey to provide an initial photometric type of the SNe and a selected sample of sources was targeted for spectroscopic observations.

In order to investigate the classification from images rather than light curves, we acquired the images from the public SDSS dataset through their platform. Our dataset contains many types of supernovas (see Table \ref{table:data} and  \citep{sdss2}).
The label of "unknown" mainly represents very sparse or poorly measured transient candidates, "variables" have signals spanning over two seasons, and "AGNs" have a spectral signature. The three other classes are supernovae of type Ia, Ib/c, and II. Among supernovae, the typing is performed from spectroscopy or from the light curve using different machine learning techniques \citep[see][]{sdss2}.  We grouped the non-Ia supernovas because our focus in this study only on the Ia type for their interest in cosmology as standard candles and also because of the small number of non-Ia with spectral signatures.
The very small class of three SLSN  bright objects has been added to the non-Ia supernovae. Figure \ref{fig:imagedata} shows an example of astronomical image time taken from the SDSS dataset.

\begin{table}[!ht]
\centering\addtolength{\tabcolsep}{10pt}\fontsize{13}{10}\selectfont

\begin{tabular}{||c c ||}

 \hline
 Object name & Count  \Tstrut\Bstrut\\ [1ex]
 \hline\hline
 \rule{0pt}{15pt}AGN & 906 \Tstrut\Bstrut\\
 \hline
 \rule{0pt}{15pt}SNIa  & 499 \Tstrut\Bstrut\\
 \hline
 \rule{0pt}{15pt}SNOther   & 89 \Tstrut\Bstrut\\
 \hline
 \rule{0pt}{15pt}Unknown  & 2009 \Tstrut\Bstrut\\
 \hline
 \rule{0pt}{15pt}Variable   & 3225 \Tstrut\Bstrut\\
 \hline
 \rule{0pt}{15pt}SNOther\_PT   & 2041 \Tstrut\Bstrut\\
 \hline
 \rule{0pt}{15pt}SNIa\_PT   & 1448 \Tstrut\Bstrut\\
 \hline
\end{tabular}
\vspace*{2mm}
\caption{Number of objects per class in the SDSS dataset. PT: Photometrically typed, which means that the SNs are not spectroscopically verified }
\label{table:data}
\end{table}

\subsection{Challenges}
 Most of the astronomical dataset suffers from a number of problems that should be dealt with before feeding it to the classification algorithm. Among difficulties contributing to the challenging nature of AITS, we can mention class imbalance (as shown in Table  \ref{table:data} of our dataset). In particular, we can clearly see that the classes we have are not balanced where the number of samples for variables is much  bigger than SNIa. This imbalance  significantly impacts machine learning models due to their higher prior probability, which means they tend to overclassify the larger class(es). As a result, instances belonging to the smaller class(es) are more likely to be misclassified than those belonging to the larger class(es). Another problem that impacts the model is missing bands. Indeed, each time an image is acquired in an AITS it is captured through one filter among a set of up to five or more channels. So, an image of a celestial object can be taken in many channels, but not necessarily at the same time. This results in missing bands for a given time of observation (see Figure \ref{fig:bandsInImage}). It is well known that the missing data negatively impacts  the performance of the model if it is not dealt with. \citet{missd} stated that an increasingly missing percentage of training data resulted in an increased testing error, which requires a solution to mitigate the impact of missing data.

\section{Methods}
\label{section:methods}
In this section, we propose a neural network based on a combination of convolution and self-attentions. The goal of the model is to handle the challenges that we mentioned previously, such as class imbalance, data sparsity, and missing observations. Figure \ref{fig:allarchi} represents the general architecture of the ConvEntion model. The model takes as its input the sequence of images that have been rearranged to embed the band information (See Section \ref{subsection:datamodeling} and Figure \ref{fig:Sepband}). The sequence  first passes through a 3DCNN to downsize its length. It allows for the reduction of the computation complexity of the model and also captures the local characteristics of the objects. The newly constructed sequence by the 3DCNN is fed to a convolutional BERT which then extracts the spatio-temporal features with high-level representation from the input. Finally, we pass the output of the convolutional BERT, which is a projection of our input into a high-level representation subspace, through a 3D max-pooling to downsample it, then it goes on to the final classifier to make the prediction. In the following subsections, we explain each component in depth.

\subsection{Data modeling}
\label{subsection:datamodeling}
\begin{figure}[ht]
  \centering{\includegraphics[width=8.9cm, height=3cm]{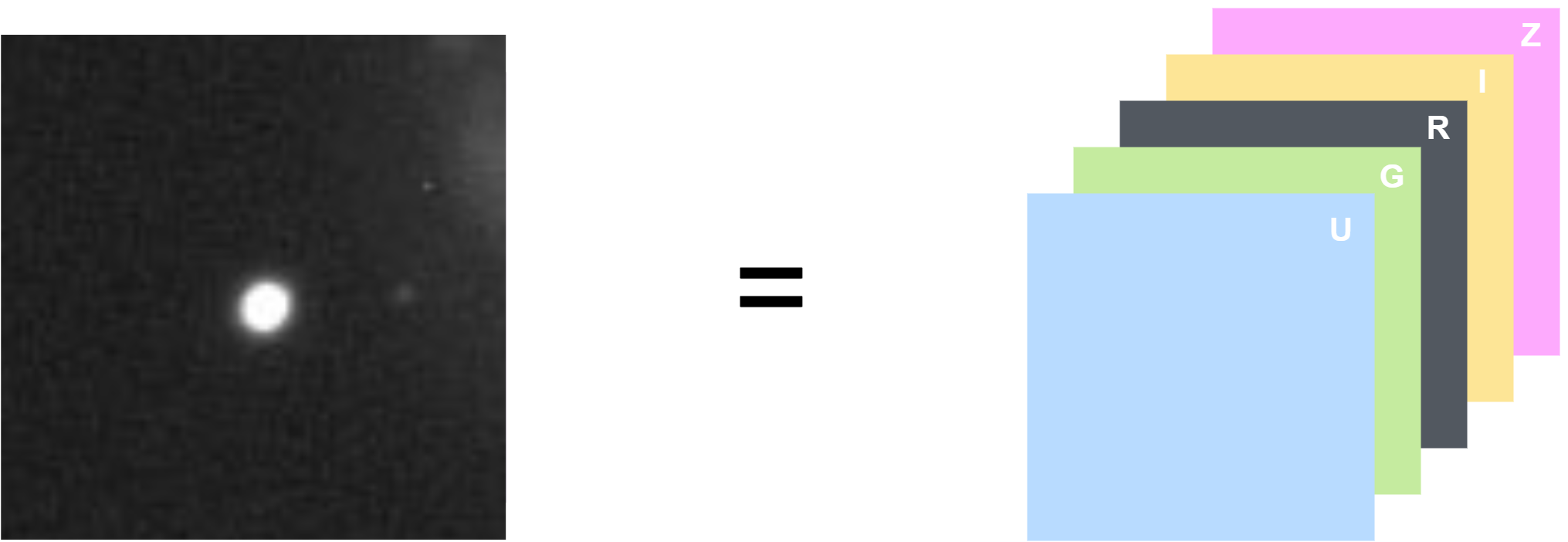}}
  \caption{Each image has five filters  (u, g, r, i, z), The black channel represents the missing observation}
  \label{fig:bandsInImage}
\end{figure}

First, we note that throughout the paper, vectors are given in bold capital letters, sizes in capital letters, and indices in lowercase.
To start with the \textbf{missing data problem}, a network dedicated to image time series is usually fed  a sequence of images $\textbf{I} \in \mathbb{R}^{H \times W \times 5}$, where $H$ and $W$ are, respectively, the height and width of the image and 5 is the number of channels representing the bands (u, g, r, i, z). However, we know, as explained earlier, some bands are missing in the dataset. To fix this issue, instead of giving the model images with empty channels, thus introducing a bias to the network, we decided to separate the channels as individual images ($\textbf{X} \in \mathbb{R}^{H \times W}$) simply by skipping the empty channels. As a consequence, the information about the type of filter, which holds a crucial value for the network to accurately discriminate between objects, is also eliminated. In an image with different channels, the order of the channels usually represents the type of filter (see Figure \ref{fig:bandsInImage}).

\begin{figure*}[ht]
  \centering{\includegraphics[width=18cm, height=9.21cm]{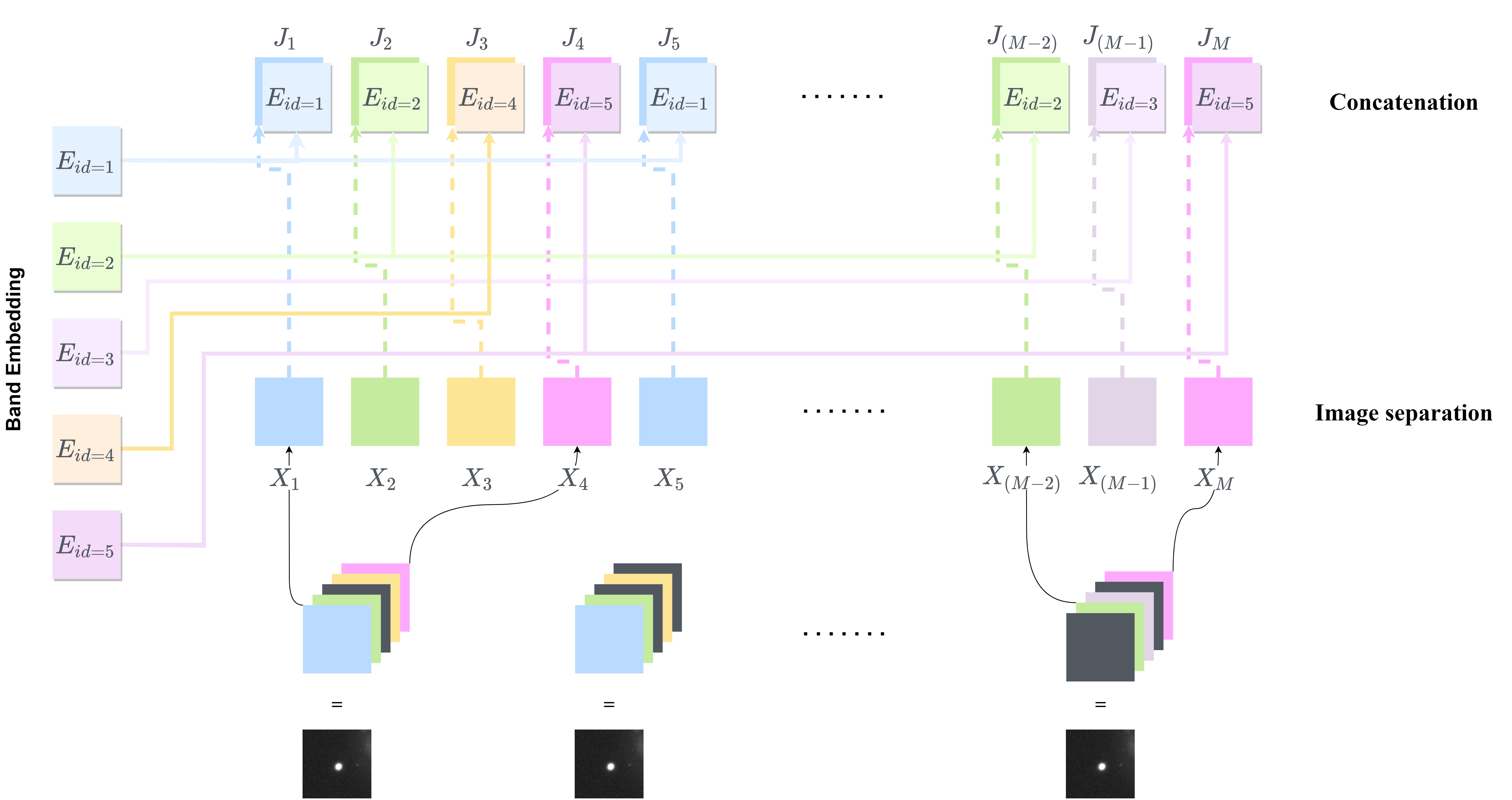}}
  \caption{ Illustration of the handling of missing information by separating the bands. The empty channels are dropped, then we concatenate each image with a 2D representation of the band used to capture the image. The band embedding contains five band representations. The black channel represents the missing observation}
  \label{fig:Sepband}
\end{figure*}
In order to preserve this valuable information, we should add the band type
to the new 2D images $\textbf{X}$. Knowing that the information
about the type of the filter is a categorical feature, thus we need to adapt it to the model 2D input representation. 
To do so, we propose using an embedding layer to encode the channel type before passing the input to the model. For each band 
(u, g, r, i, z), we assign a unique number $id \in \{1,2,3,4,5\}$. Then, an
embedding layer $BandEmbed$ converts the band type $id$, which is a categorical feature, into 2D dense representation
$E_{id}$ with $ {E_{id}} \in \mathbb{R}^{H \times W} $ (see Figure \ref{fig:Sepband}):

\begin{equation} \label{eq:filterEmm}
{E_{id}} = BandEmbed(id).
\end{equation}

The embedding layer is a fully connected layer that is reshaped to a 2D representation. The weights of $BandEmbed$ are learnable.
After getting the band embedding, we concatenate it with the new image to get our new input ${J} \in \mathbb{R}^{M \times H \times W \times 2}$ that contains the band information, where $M$ is the length of the sequence:

\begin{equation} \label{eq:filterEmm4}
{J_m} = Concat({X_m},{E_{id}}), \qquad m \in \{1,..,M\}.
\end{equation}

The problem of {\bf{class imbalance}} is one of the major challenges for any machine learning project. Some tried to solve this problem by adding a new loss function to mitigate the impact of the class imbalance. For example \citet{focal} proposed a loss function called “focal loss” which applies a modulating term to the cross-entropy loss in order to focus the learning on hard misclassified examples. However, this approach tends to produce a vanishing gradient during backpropagation  \citep{floss}. Other solutions propose the use of oversampling such as SMOTE  \citep{smote}. Those authors proposed an approach where they synthesize new samples of the minority class. However, this solution was proposed mainly for tabular data. Knowing that our data are images that contain a much higher number of features than tabular data, it appears obvious that using SMOTE may not be optimal in our case.  \citet{deepsmote} introduced a solution based on SMOTE dedicated to images called DeepSMOTE. It is aimed at generating new images for the minority class. Once again, this approach is unsuitable in our case as our dataset is not composed of images, but of a sequence of images, and it is too expensive to generate a whole new sequence. So, instead of generating a new one, we used data augmentation and weighted random sampling(WRS) \citep{wrs} on our database. We oversampled the dataset, which translates to simply altering the dataset to remove such an imbalance by increasing the number of minority classes and undersampling the data by decreasing the majority classes until we have reached a balanced dataset. In our case, the WRS was applied on a batch level. We generate balanced batches based on the probability of a sample being selected. We weighted each sample according to the inverse frequency of its label's occurrence and then sampled mini-batches from a multinomial distribution based on these weights. This means that samples with high weights are sampled more often for each mini-batch. The same sample can be reused in other mini-batches of the same epoch to increase the minority class, but with a data augmentation applied to it.  Different methods of data augmentation were used: for example, a random drop of some steps from the whole sequence to create a new one or a sequence rotation, horizontal and vertical flips, and sequence shifting, where we construct a smaller sequence from the original one  which has a bigger length than the input length of ConvEntion. 
In our implementation, we recall the dataset at every epoch, the transforms operation (augmentation) is executed and then we get different augmented data.  Using this oversampling approach has drastically improved the performance of the model. We used the function $WeightedRandomSampler$ from PyTorch \citep{pytorch} as an implementation of WRS.

\subsection{3D convolution network:}
In several deep learning applications, large transformer models have demonstrated fantastic success in obtaining state-of-the-art results. However, because the original transformer's self-attention mechanism consumes $O(M^{2})$ time and space with respect to the sequence length, $M$, training the model for a long sequence is so expensive, it causes the problem called "attention bottleneck"  \citep{b11,b10}. The problem is more severe for us because we use convolutions and 3D tensors inside the attention mechanism; for instance, the attention map is of a size $H\times W,$ so the complexity of the attention will be $O(M^{2} \times H \times W)$. Thus, our model would then be prohibitively expensive to train. In the last few years, there have been numerous proposals aimed at solving this issue. \citet{b11} demonstrated that a low-rank matrix could approximate the self-attention mechanism. They suggested a new self-attention method that minimizes total self-attention complexity. \citet{b10} presented a novel transformer architecture that uses linear space and time complexity to estimate regular (softmax) full-rank-attention Transformers with proven accuracy. However, all these propositions remain irrelevant in our case because we do not use the standard self-attention mechanism, as the convolutions make it an arduous task. So, the solution we preferred to go with is to reduce the length of the sequence before feeding it to the transformer block. Reducing the sequence must be done without losing relevant information. Thus, we propose using a 3D convolution neural network (3D CNN). A 3D CNN is an improved type version of CNN first proposed by \citet{3dcnn}, where it applies a 3D filter to the dataset and the filter moves in three directions to calculate the low-level feature representations. Their output shape is in a 3D volume space. We applied $3DCNN$ where we input the sequence $\textbf{J}$ to get the reduced new sequence $\textbf{S}$ following the equation:

\begin{equation} \label{eq:filterEmm3}
{S_n} = \text{3DCNN}({J_{(n-1)*K+1}},.., {J_{n*K}}),   \quad n \in \{1,..,N\}.
\end{equation}

\begin{figure*}[ht]
  \centering{\includegraphics[width=15cm, height=7.71cm]{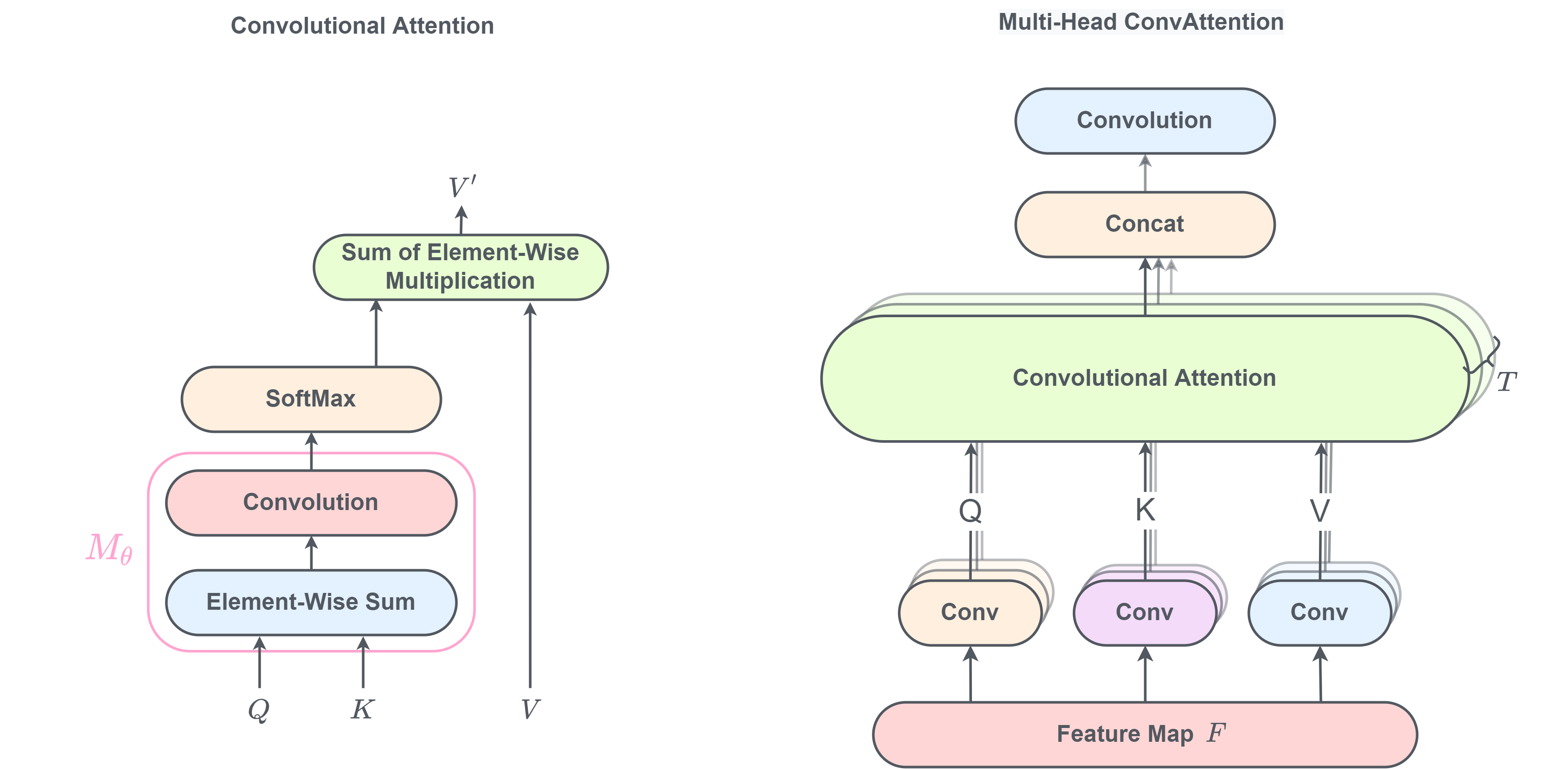}}
  \caption{Convolutional attention (left). Multi-head convolutional attention (right). To obtain the query, key, and value maps, we applied a convolution layer on the feature map obtained from 3DCNN.}
  \label{fig:attention}
\end{figure*}

We let $M$ be the length of the series, ${J}$ and we fed $K$ inputs of ${J}$ to the 3DCNN to generate one entry, ${S,}$ for our transformer. So, in the end, the new sequence, ${S,}$ will be ${S} \in \mathbb{R}^{N \times H' \times W' \times D}$, where $N = M/K$, $D$ is the number of channels and $H'$ and $W'$ are the new height and width. By using the 3DCNN, we reduced the length of the sequence by a factor of K, which also reduced the complexity of the model. The 3DCNN does not just reduce the length of the input sequence, it also captures local spatio-temporal low-level features. The 3DCNN captures these particulate features due to its focus on the local characteristics (space and time) of the sequence, while the transformer focuses on the global characteristics.
On the whole, we have reduced the computation without losing essential information that is important for classification. Table \ref{table:cnn} summarizes the architecture used inside the 3DCNN.




\begin{table}
\centering
\begin{tabular}{p{0.30\linewidth}p{0.40\linewidth}}
\hline
 Layer & Layer Parameters \\
 \hline\hline
 \rule{0pt}{15pt}Conv3d + BN3d & $11 \times 11 \times 3 \times 64, 64$ \Tstrut\Bstrut\\
 \hline
\rule{0pt}{15pt} Conv3d + BN3d & $5 \times 5 \times 3 \times 128, 128$ \Tstrut\Bstrut\\
 \hline
 \rule{0pt}{15pt}Conv3d + BN3d & $3 \times 3 \times 3 \times 64, 64$  \Tstrut\Bstrut\\
 \hline
 \rule{0pt}{15pt}Conv3d + BN3d & $3 \times 3 \times 3 \times 64, 64$ \Tstrut\Bstrut\\
\hline \\
\end{tabular}
\caption{3D CNN architecture where Conv3D is a 3D convolutional element and BN3d is a 3D batch normalization element.}
\label{table:cnn}
\end{table}

\subsection{Convolutional BERT}\label{AA1}

After getting the new output ${S}$ of the 3DCNN, it is time to feed it to what we call convolutional BERT  which stands for Convolutional Bidirectional Encoder Representations from Transformers. Transformer and self-attention have become one of the main models that revolutionize deep learning in the last few years, especially in neural language processing (NLP). Self-attention \citep{attMec}, also known as intra-attention, is an attention mechanism that connects different positions in a single sequence to compute a representation of the sequence. Here, "attention" refers to the fact that in real life, when viewing a video or listening to a song, we frequently pay more attention to certain details while paying less attention to others, based on the importance of the details. Deep learning uses a similar flow for its attention mechanism, giving particular parts of the data more focus as it is processed. Our intention in using this mechanism is for the model to focus more on the changes happening in the image sequence to better discriminate between astronomical objects. Self-attention layers are the foundation of the transformer block design. Transformers were first introduced by \citet{b2}, using model-based attention dispensing with recurrence and convolutions entirely. Their work inspired others who used the concept of transformers to achieve even better results. For example, in BERT  \citep{b8} the authors used only the encoder block by stacking many of them. Even though transformers were widely used in NLP in the last two years, people started implementing these blocks in other domains like image classification. \citet{b13} presented a model free from convolutions by using only a transformer to classify images. \citet{b23} also suggested that they are able to extract temporal characteristics using a custom neural architecture based on self-attention instead of recurrent networks. Their use was not limited to image classification; action recognition was also investigated as in \citet{b14}, where the authors used a transformer-based approach inspired by the work of  \citet{b13}. \citet{b12} did propose a new transformer where they added convolution to the attention mechanisms, making it able to apply convolutions while extracting the temporal features.

\subsubsection{Positional encoding}

Because transformers have no recurrence throughout the thumbnail sequence, some information about each thumbnail’s relative or absolute position must be injected into the feature map obtained by the 3DCNN  to inform the model about the order in the sequence. Similarly to the original transformer paper  \citep{b2}, we use positional encoding at each layer in the encoder to achieve this. The only difference is that our positional encoding is a 3D tensor, where ${P} \in \mathbb{R}^{N \times H' \times W' \times D}$. Because the positional encoding and the new feature maps have the same dimension, they can be added together. We use sine and cosine functions to encode the position  \citep{b2}:

\begin{equation} \label{eq:possin}
{P_{(n,2i)}} = sin(n/10000^{2i/D}),
\end{equation}

\begin{equation} \label{eq:poscos}
{P_{(n,2i+1)}} = cos(n/10000^{2i/D}),
\end{equation}

where $n$ denotes the position in the sequence of length, $N$, and $i$ is the channel dimension, while $D$ represent the total number of channel gotten by the
3DCNN. The sinusoidal positional encoding is chosen to make it easy for the model to learn to attend to relative positions. To get the new input for the convolutional BERT, we conducted an element-wise addition between the positional encoding  and the feature maps obtained from 3DCNN to obtain the new tensor ${F} \in \mathbb{R}^{N \times H' \times W' \times D}$:

\begin{equation} \label{eq:possum}
{F_{n}} = {S_{n}} + {P_{n}}, \qquad n \in \{1,..,N\}.
\end{equation}

In this study, we only used information about the position of the image in a sequence. While the observation date could be used as an alternative to the position, this would require adjusting the positional encoding function. Our experiments on the SDSS dataset did not reveal any improvement in the model when using the observation date, as opposed to just using the position. This can be understood because we do the training and the test with the same observation sequence and the network can therefore learn this sequence. On the other hand, not incorporating any information regarding the order of the sequence greatly degraded the performance of the model. As a result, we ultimately chose to use only the position in our model (see Section \ref{section:experiments:results} for a discussion).

The newly obtained sequence $F$ is fed to a multi-head convolutional attention, which is an improved self-attention that has convolution. Then the multi-head convolutional attention is followed by the second component which is a tiny feed-forward network (FFN) that has convolutions applied to every attention map. Its primary purpose is to transform the attention map into a form acceptable by the next convolutional BERT layer, with the FFN consisting of two convolutional layers with ReLU activation in between.

\subsubsection{Multi-head convolutional self-attention}\label{AA}

For this process, we used the model proposed by  \citet{b12}, with a few modifications where we replaced the last linear layer with a convolution layer. We believe that convolution in self-attention is better than the dot product between the query and the key because the convolution will accurately calculate the similarity, especially when we have 3D feature maps. A query map and a set made up of a pair of key maps and value maps that are encoded to an output using convolutional self-attention. The query map, key maps, value maps, and output are all 3D tensors. Figure \ref{fig:attention} represent the general architecture of the multi-head ConvAttention.


We used a  convolution layer to generate the attention model’s query, value, and key. The input to the attention model is ${F} \in \mathbb{R}^{N \times H' \times W' \times D}$. We pass each map through a convolution layer to get $\{{Q},{K},{V}\} \in \mathbb{R}^{N \times  H' \times W' \times D'}$, where $D' = D/T$ and $T$ represent the number of attention heads. Then we appled a subnetwork, ${M_{\theta}}$, on the query and the key maps, which consists of an element-wise sum of the query and the key maps followed by another convolution layer to generate our attention map ${H_{(n,m)}} \in \mathbb{R}^{H' \times W' \times 1 }$:

\begin{equation} \label{eq:attention}
{H_{(n,m)}} = M_{\theta}({Q_{n}},{K_{m}}),   \qquad n, m \in \{1,..,N\}.  
\end{equation}

 After getting all the map attentions, ${H_{n}} = \{{H_{(n,1)}},{H_{(n,2)}},....,{H_{(n,N)}}\}$, where ${H_{n}}\in \mathbb{R}^{H' \times W' \times N}$, we applied a softmax operation along the third dimension of size, $N$. Then we conducted an element-wise product between the attention map and the value map following the equation:
 

\begin{equation} \label{eq:soft}
V'_{n} = \sum\limits_{m=1}^N SoftMax({H_{n}})_{(n,m)}V_{m}.
\end{equation}


We concatenated the new value representation, $V'_{n}$, obtained from the different attention heads. The multi-head attention is used to attend to input from various representation subspaces jointly:

\begin{equation} \label{eq:conca}
MultiHead(Q,K,V) = Concat(V'_{n_{1}} ,...., V'_{n_{T}}).
\end{equation}
\\
Finally, we applied a convolution layer for merging the output of the multi-head and obtaining a high-level representation that groups all the heads. At the end of the network, we pass the encoded sequence to 3D max-pooling and finally to the classifier to make a prediction.

\subsection{Evaluation metrics}

 Accuracy is the probability that an object will be correctly classified. It is defined as the sum of the true positives plus true negatives divided by the total number of individuals tested:

\begin{equation} \label{eq:accu}
Accuracy = \frac{TP+TN}{TP+TN+FP+FN}
,\end{equation}

where TP, TN, FP, and FN are, respectively, the true positive, true negative, false positive, and false negative.

The F1 score is a classification accuracy metric that combines precision and recall. It is a suitable measure of models tested with imbalanced datasets:\ 
\begin{equation} \label{eq:Precision}
Precision = \frac{TP}{TP+FP}
,\end{equation}

\begin{equation} \label{eq:Recall}
Recall = \frac{TP}{TP+FN}
,\end{equation}

\begin{equation} \label{eq:F1}
F1 = 2 \times \frac{Precision \times Recall}{Precision+Recall}
.\end{equation}

\begin{table*}[hbt!]
\centering
\begin{tabular}{p{0.20\linewidth}p{0.15\linewidth}p{0.15\linewidth}p{0.10\linewidth}p{0.10\linewidth}p{0.10\linewidth}}
\hline \\
Model &Bands & Type of data &  Accuracy   &  F1 Score & Num params \Tstrut\Bstrut\\
\hline
\rule{0pt}{15pt}ConvEntion (Ours) & ugriz & Images & \textbf{79.83} & \textbf{70.62}  & 1.253M \\
\rule{0pt}{15pt}CNN+GRU    \citep{b21}  & ugriz & Images & 66.39 & 63.22  & 1.993M \\
\rule{0pt}{15pt}ConvEntion (Ours) & g & Images & 76.89 & 63.20 & 1.253M\\
\rule{0pt}{15pt}CNN+GRU    \citep{b21}  & g & Images & 63.67 & 61.00  & 1.992M \\
\rule{0pt}{15pt}CNN+LSTM  \citep{b20} & ugriz & Images & 64.08 & 60.65 & 2.190M \\
\rule{0pt}{15pt}CNN+LSTM  \citep{b20} & g & Images & 63.00 & 60.00 & 2.189M \Bstrut\\
\hline
\hline
\rule{0pt}{15pt}SuperNNova (Bayes)  \citep{b18} & ugriz &  Light curves & 65.54 & 55.40  & - \Tstrut\\
\rule{0pt}{15pt}SITS-BERT   \citep{sbert} & ugriz &  Light curves & 67.43 & 51.60  & 0.596M \\
\rule{0pt}{15pt}SCONE (CNN)  \citep{Scone}  & ugriz &  Light curves & 62.57 & 50.43  & 22.2K \\
\rule{0pt}{15pt}SuperNNova (RNN)   \citep{b18} & ugriz & Light curves & 56.30 & 42.60  & - \\
\rule{0pt}{15pt}LSTM  & ugriz &  Light curves & 55.24 & 40.33  & 60K \\

\hline \\
\end{tabular}

\caption{Performance comparison in terms of average F1 score and the average of the accuracy of five folds of cross-validation}. This table includes only experiments on a dataset with four classes. 
\label{table:bench}
\end{table*}

\begin{table*}
\centering
\begin{tabular}{p{0.30\linewidth}p{0.10\linewidth}p{0.20\linewidth}p{0.20\linewidth}}
\hline
Model &Bands &   Accuracy   &  F1 Score \\
\hline
\rule{0pt}{15pt}ConvEntion (Ours) & ugriz & \textbf{83.90} & \textbf{75.77}\\
\rule{0pt}{15pt}ConvEntion (Ours) & g &  79.47 & 72.38\\
\rule{0pt}{15pt}CNN+GRU   \citep{b21} & g &  74.84 & 68.95 \\
\rule{0pt}{15pt}CNN+LSTM  \citep{b20} & g & 73.94 & 67.29 \\
\hline \\
\end{tabular}
\caption{Performance comparison in terms of average F1 score and the average of the accuracy of five folds of cross-validation. This table includes only experiments on a dataset with three classes.}
\label{table:ressum3c}
\end{table*}

\section{Experiments}
\label{section:experiments}
\subsection{Implementation details}
The supernovae in our data are not all spectroscopically confirmed, which means that the unconfirmed ones  might contain some misclassified objects due to errors from the photometric typing. The model may not generalize due to this data bias. To ensure that our model performs a generalization only on  spectroscopically confirmed data, we split up the training process into two steps. We divided the data into two datasets. The first one contains only the photometrically typed data and the second contains spectroscopically confirmed data. We trained the model at first with the  photometrically typed data, then we used transfer learning to fine-tune the model on only spectroscopically confirmed data (Table \ref{table:datapart} summarizes the partition of the data). The models are trained using cross-validation of five folds and three ensembles in each fold. All the architectures presented in this paper follow this same process and are implemented using PyTorch \citep{pytorch}.

\begin{table}[hbt!]
\centering
\begin{tabular}{p{0.30\linewidth}p{0.15\linewidth}p{0.15\linewidth}p{0.15\linewidth}}
\hline
Class &  Train &   Fine-tune   &  Test \Tstrut\Bstrut\\
\hline
\rule{0pt}{15pt}AGN  & 362 & 362 & 182 \\
\rule{0pt}{15pt}SNIa  & 1448 &  400 & 99\\
\rule{0pt}{15pt}Variable & 1290 &  1290 & 645 \\
\rule{0pt}{15pt}SNOther & 2041 & 72 & 17 \\
\hline \\
\end{tabular}
\caption{ Count of every object in a dataset of each step in training protocol. Train contains only photometrically typed data, "fine-tune" and "test: contain only spectroscopically confirmed data}
\label{table:datapart}
\end{table}

We performed an extensive hyperparameter tuning of over 20 models to specify the best hyperparameters for our architecture, which contains 1.3 Million parameters. We conducted a hyperparameter optimization using only a non-confirmed dataset with different parameters, such as sequence length, $M$, learning rate, $lr$, 3DCCN sub-sequence length, $K$, classifier layers' size, number of ConvBERT layers, $L$, number of Multi-head ConvAttention, $T$, batch size, and dropout.
We used an Adam optimizer  \citep{adam}, with a value of  the learning rate of $10^{-3}$, and we trained the model with cross-entropy loss and a  dropout of $0.3$. Hyperparameter tuning involves the number of images K that feed the 3DCNN and the maximum length of the sequence. The best values were $K=3$ and $M=99,$ which means the number of sequences for the convolutional BERT is $N=33$. The batch size was 128 sequences which we ran over 100 epochs. We chose the number of convolutional BERT layers to be $L=2$ and the number of attention heads $T=4$. Also, the images were normalized band-wise, as each band has different characteristics. We used only four classes (AGN, SNIa, Variable, SNOther) to train all the models.
The class marked as "unknown" has not been considered in the study. It corresponds to noisy or very sparse data. It can easily be tagged from sparsity or noise in the image metrics and we do not expect any improvement in the classification if such objects are added to the training.
We trained all models with 4 GPUs GeForce RTX 2080 Ti, Each model takes about three hours to complete training. The implementation will be released upon publication in our Github page \footnote{https://github.com/DaBihy/ConvEntion}.

\subsection{Results}
\label{section:experiments:results}
This section provides studies on SDSS comparing the accuracy and F1 score of our proposed solution with other works.
Table \ref{table:bench} summarizes the result of \textbf{different models from different deep learning areas} to diversify our benchmark as it contains RNN architectures (SuperNNova, LSTM), CNN-based models such as SCONE, Hybrid models that have CNN and RNN such as  \citep{b20} and   \citep{b21}, and, finally, a transformer-based model. Also, we compared the result using two types of datasets: first, the image dataset and, second, the same dataset object but with the light curves; the goal is to highlight the advantage of using images instead of light curves. Moreover, the different works mentioned in Table \ref{table:bench} were initially proposed for different datasets with different classes and training protocols. Hence, the results do not reflect the quality of these works on other datasets. The goal of the comparison is to give visibility into the performance of our model from a deep learning standpoint and the importance of using image time series from an astronomy perspective.

Overall, our model  ConvEntion obtains the highest accuracy of 79.83\% and F1 score of 70.62\%, 13 points higher in accuracy than the best results on images by  \citep{b21} and 12 points higher in accuracy than the best model using light curves. This confirms the advantage of using images over light curves.
This advantage can be explained by the fact that the image contains more information than a single value of flux in a light curve. Hence, a model can learn robustly with the existence of more high-level feature maps. Also, ConvEntion performed better compared to the other image-based models, such as  \citet{b20}. Additionally, transformers give a remarkable computational advantage because transformers avoid recursion and allow for parallel computation, thus reducing the training time. Our model took only three hours to train, compared to other image-based models which took five hours of training on our GPUs. Our model achieved better results using fewer parameters, compared to the other models trained on image sequences. The main benefit of using a transformer is that it reduces the drop in performance due to long dependencies. Transformers do not rely on past hidden states to capture dependencies with previous features such as RNNs. They instead process a sequence as a whole. Therefore, there is no risk of losing past information. Also, the integration of a spatio-temporal feature extraction helped in getting a better high-level representation of the sequence, in comparison to separating the spatial features from the temporal ones. The two types of features have correlations that may help the model to better discriminate between objects. We can also highlight the importance of separating the band to mitigate the impact of missing observations. Our model performed well, in comparison to that of  \citet{b21} which uses multiple bands, which shows that separating the bands and adding band embedding works better than feeding the network with empty bands.

\begin{figure}[hbt!]
  \centering{\includegraphics[width=9cm, height=9cm]{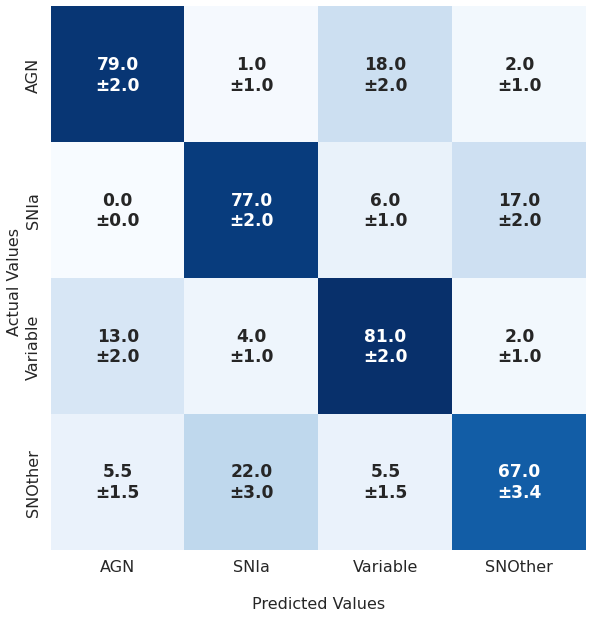}}
  \caption{Confusion matrix showing the average accuracy and standard deviation of the predictions generated by ConvEntion over cross-validation of five folds on test data.} 
  \label{fig:cm}
 
\end{figure}


 \begin{figure}[hbt!]
  \centering{\includegraphics[width=9cm, height=9cm]{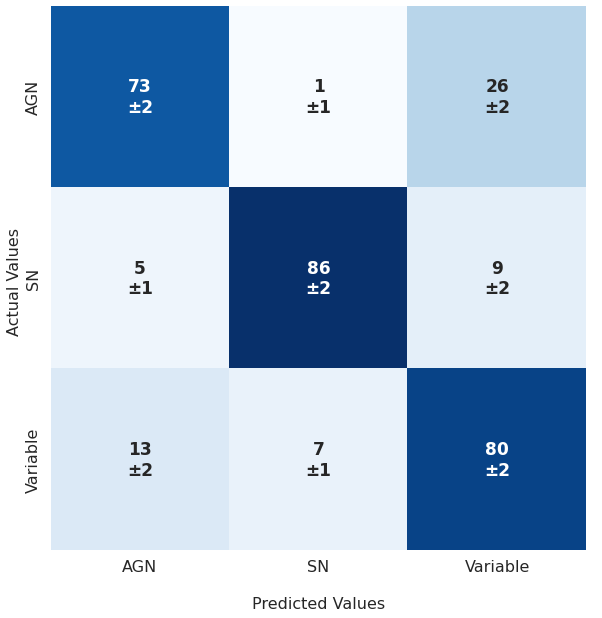}}
  \caption{Confusion matrix of three classes showing the average accuracy and standard deviation of the predictions generated by ConvEntion over cross-validation of five folds on test data.}
  \label{fig:mc3}
\end{figure}

In  the study of \citet{b20}, the authors trained their model on a dataset that only has  a "g" band and they noted that the model can be adapted to classify the image sequence combining information using multiple bands. For the sake of comparison, we trained the image models with all the bands "ugriz" at first and then with only one  "g" band. Our model achieved an accuracy of 76.89\% and 63.20\% in the F1 score using one band ("g")  which dropped 7\%  in comparison to using multiple bands. Meanwhile,  \citet{b20} achieved 63\% in accuracy and 60\% in their F1 score.  This shows that our model is more efficient 
when using multiple bands. This also highlights the impact of band separation to mitigate the impact of the missing observations.

Figure \ref{fig:cm} illustrates the obtained confusion matrix by ConvEntion and it shows that the model has well classified the supernovas. Most of the misclassified SNIa are associated with SNOther and vice versa, which is not a serious error. This is even an expected behavior, especially since all types of supernovas share a lot of similarities which may confuse the model. Additionally, with a small dataset like ours, it is normal to have such behavior because the model does not have enough samples to totally discriminate among objects. Meanwhile, variables were the best-classified class in our dataset, with just a bit of confusion with the AGN; this misclassification between AGN and variable can be explained by the class imbalance in our dataset based on the knowledge that the number of variables is higher than in the other classes.

Table \ref{table:ressum3c} summarizes the results of different models trained only on three classes (AGN, SN, Variable), where classes SNIa and SNOther are combined into a single class. The goal of this experiment is to see the behavior of our model in discriminating between transient and non-transient objects. We got the best results with an accuracy of 83.90\%  with an F1 Score of 75.77\%. The model was able to classify the SN accurately, with a score of 86\% (as shown in Figure \ref{fig:mc3}).

The model is able to effectively process a given survey without any loss in performance and without the requirement of providing it with the time information for each image. However, when there is a covariate shift, or a mismatch, between the training set and the test set  as when using a different dataset with a different observation sequence), incorporating the time information can improve the results. This experimental finding will be further studied and reported in future work using other datasets.

\section{Conclusion}
\label{section:conclusion}
In this work, we present a method for efficient astronomical image time series classification that is entirely based on the combination of convolutional networks and  transformers. Inspired by action recognition and satellite image time series classification, we propose a model ConvEntion that utilizes convolutions and transformers jointly to capture complex spatio-temporal dependencies between distinct steps, leading to accurate predictions based on different observations of an object. The accuracy of our model is better with a high margin of 13\%, in comparison to state-of-the-art methods using image data -- and even better compared to approaches using light curves. 

Our model achieves good results on the SDSS dataset, while also being faster thanks to using fewer parameters and parallel computational processes, making it a good candidate for latency-sensitive applications such as the real-time thumbnail classifier of astronomical events.
Meanwhile, our benchmark stands as clear evidence of the importance of images in the domain of astronomy. Indeed, the images contain more information than the normal light curves, even if they present more difficulties. In the future, we plan to scale up ConvEntion using self-supervised learning to investigate whether the model can generalize even better. With a large amount of unlabeled data in astronomy, we believe that the next step to advance AITS classification is creating self-supervised models.

\begin{acknowledgements}
     This work has been carried out thanks to the support of the DEEPDIP ANR
project (ANR-19-CE31-0023). This work makes use of Sloan Digital Sky Survey (SDSS) data. Funding for SDSS-III has been provided by the Alfred P. Sloan Foundation, the Participating Institutions, the National Science Foundation, and the U.S. Department of Energy Office of Science. The SDSS-III website is \url{http://www.sdss3.org/}. SDSS-III is managed by the Astrophysical Research Consortium for the Participating Institutions of the SDSS-III Collaboration including the University of Arizona, the Brazilian Participation Group, Brookhaven National Laboratory, Carnegie Mellon University, University of Florida, the French Participation Group, the German Participation Group, Harvard University, the Instituto de Astrofisica de Canarias, the Michigan State/Notre Dame/JINA Participation Group, Johns Hopkins University, Lawrence Berkeley National Laboratory, Max Planck Institute for Astrophysics, Max Planck Institute for Extraterrestrial Physics, New Mexico State University, New York University, Ohio State University, Pennsylvania State University, University of Portsmouth, Princeton University, the Spanish Participation Group, University of Tokyo, University of Utah, Vanderbilt University, University of Virginia, University of Washington, and Yale University.
\end{acknowledgements}

%
%
\bibliographystyle{aa}
\bibliography{biblio}

\end{document}